%% file: arxiv.tex
\documentclass{lipics-v2021}
\usepackage{amssymb,amsfonts,amsmath,amsthm,amscd,mathtools,proof}
\input{symbol}
\input{unit}

\input{font}
\input{mathenv}
\input{color}
\newcommand{\nextt}{\\ \zero \quad}
\counterwithout{equation}{section}
\counterwithout{equation}{subsection}

\renewcommand{\bit}[1]{{\small{\textbf{\textit{#1}}}}}	
\newcommand{\flp}{\bxsf{FL}$\! ^+$}
\newcommand{\esl}{\flp}
\newcommand{\fol}{\bxsf{FL}}
\newcommand{\fl}{\bxsf{FL}}
\newcommand{\cl}{\bxbf{Cl}}
\renewcommand{\hl}{\bxsf{HL}}
\newcommand{\sol}{\bxsf{SOL}}

\newcommand{\ccup}{\,{\pmb{\cup}}}

\newcommand{\vv}{{\small{\grS}}}

\begin{document}
\hideLIPIcs
\nolinenumbers
\title{A foundational characterization of Hoare Logic}
\author{Daniel Leivant}{Computer Science Department, Indiana University Bloomington, USA}{}{}{}{}

\authorrunning{D. Leivant}

\Copyright{Daniel Leivant}

\keywords{Hoare Logic, Proof theoretic equivalence, first-order comprehension, second-order logic}

\ccsdesc[500]{Theory of computation~Logic and verification}

\maketitle

\begin{abstract}
We show that a partial-correctness assertion about an iterative program is provable in Hoare Logic iff it is provable in standard second-order logic with comprehension restricted to first-order predicates.  This equivalence was claimed twice in the past, both with faulty proofs, and seems to be the first foundational characterization of Hoare's Logic.
\end{abstract}

\section{Introduction}

Hoare-style formalisms are arguably the predominant paradigm of imperative 
program veri\-fi\-cation.
Their building blocks are partial-correctness assertions (pca's)
i.e.\ phrases \SM{\grf\,\plbk \gra \prbk\, \grq}
with \sms{\gra} a program and \sms{\grf,\grq} first-order formulas (ff's).
This asserts that \sms{\gra}'s output satisfies \sms{\grq} 
provided its input satisfies \sm{\grf}.
A pca is satisfied vacuously when the input yields no output,
so nothing is being said about program termination (``totality''), 
hence the qualifier ``partial.''

Since the set of pca's that are Hoare-provable is RE,
whereas the set of pca's valid in all structures 
is not \cite{GergelyS80}, Hoare Logic is not semantically complete,
unlike first-order logic (\fl).
Are then foundational properties that do characterize 
the provable pca's?

A foundational characterization of Hoare Logic,
proposed in \cite{Leivant85}, is to match Hoare Logic
with an innocuous extension \flp\ of first-order logic,
which admits relational quantification but restricts 
the Comprehension Rule to first-order definable predicates. 
\flp\ combines the expressive power of relational 
quantification with the ontological and deductive 
clarity of first-order logic. 
Unfortunately, the proof sketched in \cite{Leivant85} relies on a faulty
formal definition of program semantics.

We show here that \hl\ for iterative programs is indeed
characterized by provability in second-order logic with first-order 
comprehension.
It is a proof-theoretic
matching two forms of provability, thereby solidifying a long-held tenet
that Hoare Logic is syntactic sugar for first-order reasoning 
about pca's.
The characterization holds also for
the constructive (i.e.\ intuitionistic) readings of the
two formalisms.

\section{Related works}

Note from the outset that our characterization is unrelated to the extensive 
literature on second-order methods in program verification. 
In fact, it is the proof-theoretic weakness of \flp\ that makes
it relevant here.

An obvious question is whether Cook's relative completeness
\cite{Cook78} is related.
Relative Completeness 
epitomize the ``one structure at a time'' approach,
and as such fails to characterize \hl:
if \sms{\grp} is any pca that is valid but unprovable in \hl\
then extending \hl\ with \sms{\grp} yields a proper extension of \hl\
that is also sound and relatively complete.
Such \sms{\grp} must exist since \hl\ is RE while the set of 
valid pca's is not.

Blass and Gurevich observed that expressiveness is guaranteed if 
existential fixpoints are permissible in pre- and post-conditions
\cite{BlassG01}.
This trivializes the expressiveness condition of
relative completeness at the cost of a second-order extension
of the underlying language.
Unfortunately, this still does not provide a characterization
of \hl: the set of valid pca's, now with existential-fixpoint permissible
in the pre- and post-conditions, fails to be RE (otherwise
the set of first-order pca's would be RE as well).

Makowsky and Sain \cite{Makowsky-Sain-tcs89} proposed a model-theoretic
proof for the proposal in \cite{Leivant85}.
Unfortunately it missed the mark.
In the first place it refers to a different logic, 
one where relational quantifiers 
range precisely over the first-order definable predicates.
Lindstrom showed that the valid formulas of this logic form a Turing-complete 
\sm{\grP_1^1} set \cite{Lindstrom73}, let alone axiomatizable,
and its relevance to program verification is therefore questionable.
Moreover, \cite{Makowsky-Sain-tcs89} based the completeness
of \hl\ for their logic on \cite{Csirmaz85},
which refers only to single-loop programs,
disregarding the central role of modular program development 
and verification that underlies the very rationale of Hoare Logic.

\section{Hoare Logic for iterative programs}

\subsection{Basic conventions}


We refer to the usual notion of a \bit{vocabulary,} consisting
of function- and relation-identifiers, each assigned 
an arity \sm{\geq 0}.
We posit a 
set \SMS{V = \{ v_1, v_2 \ldots \}} of (first-order) \bit{variables}.
The \sms{\grS}-\bit{terms} 
are generated from variables and function identifiers as usual.
Thus every nullary function-identifier (a ``constant'') is a term. 
The \bit{atomic-formulas} are the
formal equations between terms, and the
phrases \sms{\ttQ(\bmt_1 ,\ldots,\bft_k)} where \sms{\ttQ} is a
\sm{k}-ary relational identifier in the vocabulary.
The \bit{first-order formulas} (ff's) are generated from atomic-formulas
using connectives and quantifiers, as usual.
If \sms{E} is a term or a formula then \sms{\grh} is \bit{adequate} 
for \sms{E} if 
it is defined for all variables free in \sm{E}.  For such \sms{\grh}
we write \sms{\grh_E} for the restriction of \sms{\grh} to variables free in \sm{E}.
For distinct variables \SMS{x_1 \ldots x_k} and 
elements \sms{a_1 \ldots a_k \in |\calS|}, we write
\SMS{[\vec{a}/\vec{x}]} for the valuation that maps \sms{x_i} to \sm{v_i \in |\calS|}
(\mm{i=1\ldots k}).
\sms{\grh\cdot \grh'} is the valuation that overrides \sms{\grh} by \sms{\grh'}:
\sm{(\grh\cdot \grh')(x) = \grh'(x)} if defined, and \sm{=\grh(x)} if not.
We also use brackets for syntactic substitution:
If \sms{E} is a term or a formula with free variables
\SMS{x_1 \ldots   x_k} then
\SMS{E[\bmt_1, \ldots, \bmt_k \,/ \, x_1, \ldots , x_k ]} 
is the result of simultaneously substituting in \sms{E}
the terms \sms{\bmt_i} for free occurrences of \sm{x_i} ($i=1\ldots k$).
When \sms{\vec{x}} is implicit we write \sms{E[\bmt_1, \ldots, \bmt_k]}.

A \vv-\bit{structure} \sms{\calS} consists of
a set (\ms{\calS}'s \bit{universe,} denoted \sms{|\calS|});
an interpretation of each \sms{k}-ary function-identifier
\sms{\ttf} as a \sm{k}-ary function 
\SM{\ttf\!_{\scriptcalS}: \; |\calS|^k \pa |\calS|}; and
an interpretation of each \sms{k}-ary relation-identifier
\sms{\ttQ} as a \sm{k}-ary relation \SM{\ttQ_{\scriptcalS} \subseteq |\calS|^k}.
A \bit{valuation (in $\calS$)}
is a function \SM{\grh: \; V \pa |\calS|}.
We write \sms{\bxit{Val}_{\scriptcalS}} for the set of such valuations.
An \bit{interpretation} is a pair \sms{(\calS,\grh)} where \sms{\grh}
is a valuation into \sm{\calS}.
\ms{\bmt_\eta} denotes the value of the \mm{\grS}-term 
\sms{\bmt} under that interpretation (we leave \sms{\calS} implicit).
\SMS{\calS,\grh \models \grf}
states that
\sms{\grh} is adequate for \sms{\grf} 
and \sms{\grf} is true in \sms{(\calS,\grh)}.

\subsection{Programs}

%

We refer to the constructs of
Pratt and Segerberg for regular programs \cite{Pratt76,Segerberg77,Harel-DL}, 
for which Hoare Logic is particularly suited. 

\bdfn
\bit{Programs} \sms{\gra} are generated 
inductively by the following clauses. 
If \sms{\grb} and \sms{\grg} are programs
then so are the following.

\noindent
%
\ph\bit{Assignment.}
\sms{x_1,\ldots, x_k := \bmt_1,\ldots , \bmt_k}, where 
\SMS{x_1\ldots x_k \in V} are distinct and
\sms{\bmt_1\ldots \bmt_k} are \vv-terms.\\[1mm]
%
\ph\bit{Test.} \sms{?\grf} where \sms{\grf} is a \vv-formula.\\[1mm]
%
%
\ph\bit{Composition.}
\sms{\grb\,\comp\, \grg}\\[1mm]
%
\ph\bit{Union.} \sms{\grb\ccup\, \grg}  \\[1mm]
%
\ph\bit{Iteration.} \sms{\grb^\star}.

\edfn


Traditional constructs are definable in terms of the operations above:
\zero \bxsf{if} \mm{\grf} \bxsf{then} \mm{\gra} \bxsf{else} \mm{\grb} 
is definable as \sm{(?\grf \comp \gra) \, \ccup \, (?\neg\grf \comp \grb)};
and \bxsf{while} \mm{\grf} \bxsf{do} \MS{\gra} as
\sms{(?\grf \comp \gra)^*(?\neg \grf)}.

%

We outline the denotational semantics of programs without dwelling on
the role of input and output, which can be easily incorporated.  

Since we allow nondeterministic programs the semantics of a program is
a \bit{mapping}, i.e.\ a binary relation mascarading as a multi-function.
We use the notation \sms{f: \,D \rA R} for a mapping that to each
input \sms{a \in D} returns a (possibly empty, finite, or infinite)
set \sm{B \subseteq R} of allowed outputs.

\bdfnl{dfn:semantics}
The \bit{semantics} of a program \sms{\gra} with variables \sms{x_1 \ldots x_k} 
in a \vv-structure \sms{\calS},
denoted \sms{\bar{\gra}_{\scriptcalS}} (subscript omitted when not needed),
is a mapping of type \sm{\bxit{Val}_{\scriptcalS} \rA \bxit{Val}_{\scriptcalS}}.
The inductive definition is by cases for \sm{\gra},
where we write \sms{\grh \ara{\alpha} \grh'} for \sm{\grh' \in \bar{\gra}(\grh)}:
\\[1mm]
%
%
%
\ph \bit{Assignment:} \sms{x_1,\ldots, x_k \,:=\, \bmt_1,\ldots , \bmt_k}.\\
\sms{\grh \ara{\alpha} \grh'} iff \sms{\grh'(x_i) = (\bmt_i)_{\eta}}
and \sms{\grh'(y) = \grh(y)} for all other variables \sm{y}.\\[2mm]
%
\ph \bit{Test:} \sms{?\grf}.\nextt  
\MS{\grh \ara{\alpha} \grh'} iff
\SM{\eta' = \eta} and \sm{\calS, \eta \models \grf}.\\[2mm]
%
%
\ph \bit{Composition:} \sms{\grb \comp \grg}.\nextt
\MS{\grh \ara{\alpha} \grh'} iff
\SMS{\grh \ara{\beta} \grh'' \ara{\gamma} \grh'} for some \sm{\grh''}.\\[2mm]
\ph \bit{Union:} \sms{\grb \ccup \, \grg}.\nextt
\MS{\grh \ara{\alpha} \grh'} iff 
\SMS{\grh \ara{\beta} \grh'} or \SM{\grh \ara{\gamma} \grh'}.\\[2mm]
\ph \bit{Iteration:} \sms{\beta^\star}.\nextt
\MS{\grh \ara{\alpha} \grh'} iff
\SMS{\grh = \grh_0 \ara{\beta} \grh_1 \ara{\beta} \cdots \ara{\beta} \grh_k = \grh'} 
for some \SMS{k\geq 0} and \SM{\grh_1\ldots \grh_k \in \bxit{Val}_{\scriptcalS}}.
\edfn


\subsection{Hoare's Logic for iterative programs}

\bdfn
A \bit{partial-correctness} \vv\bit{-assertion (pca)} is a phrase of the form
\bxrm{$\grf\, \plbk \, \gra \, \prbk \,\grq$}, where \sms{\grf} and \sms{\grq} are
\vv-formulas and \sms{\gra} is a \vv-program. 
Such a pca is \bit{true}
in an interpretation \sms{(\calS,\grh)} iff for all \sms{\grh' \in \bxit{Val}_{\scriptcalS}},
if \sms{\calS,\grh \models \grf} and \sms{\grh \ara{\alpha} \grh'}
then \sms{\calS,\grh' \models \grq}.

As usual, we refer to \sms{\grf} above as the pca's \bit{pre-condition} and
to \sms{\grq} as its \bit{post-condition}.
\edfn

\bdfn
Hoare Logic \hl\ for iterative programs
is the deductive system whose phrases are \vv-pca's and \vv-formulas,
and whose inference rules are the rules of first-order logic as
they apply to \vv-formulas, and the following rules for deriving pca's.
\edfn\zero\\
\begin{tabular}[t]{lcl}
\begin{tabular}[t]{ll}
\ph \bit{Assignment:} &
	\mm{\grf[\bmt/x] \; \plbk\, x := \!\bmt \,\prbk\; \grf}\\[3mm]
%
%
\ph \bit{Test:} &
	$\begin{array}[m]{l}
		\infer{\grf \; \plbk\, ?\,\grx \, \prbk\; \grq}
	  {\grf \wedge \grx \ra \grq}
	\end{array}$\zero\\[5mm]
\ph \bit{Composition:} &
	$\begin{array}[m]{l}
	\infer{\grf \; \plbk \,\grb\comp \grg \,\prbk\; \grq}
	{\grf \; \plbk\,\grb\,\prbk\; \grch & 
				\grch \; \plbk\,\grg\,\prbk\; \grq}
	\end{array}$
\end{tabular} 
&&
\begin{tabular}[t]{ll}
\ph \bit{Union:} &
	$\begin{array}[m]{l}
	\infer{\grf \; \plbk\,\grb \ccup \grg \,\prbk\; \grq}
	{\grf \; \plbk\, \grb\,\prbk\; \grq & \grf \; \plbk\,\grg\,\prbk\; \grq}
	\end{array}$\zero\\[5mm]
%
%
\ph \bit{Iteration:} &
	$\begin{array}[m]{l}
	\infer{\grf \; \plbk \, \gra^* \, \prbk \;  \grf}
		{\grf \; \plbk\, \gra \,\prbk \; \grf}
	\end{array}$\zero\\[5mm]
\ph \bit{Consequence:} &
	$\begin{array}[m]{l}
	\infer{\grf \;\plbk\,\gra \, \prbk\; \grq}
  	{\grf \ra \grf' & \grf' \; \plbk\,\gra \,\prbk \; \grq' &\grq' \ra \grq}
	\end{array}$
\end{tabular}
\end{tabular}

\medskip

A pca \sms{\grp} 
is \bit{provable} in \hl\ if there is a derivation of \sms{\grp}
using \fol\ to derive formulas and the rules above to derive pca's. \qed

\medskip

\bthm\label{thm:sound-rules}
The rules of \hl\ are semantically sound:
if the premise(s) of a rule are true in \sms{(\calS,\grh)} for all \sms{\grh}
then so is the conclusion.
Consequently, if a pca is derivable then it is valid, i.e.\ true in
all interpretations.
\ethm

\section{Enhanced first-order logic}

\subsection{Second-order formulas}

The semantics of programming languages
has been described formally using higher-order constructs
multiple times over the decades. 
Towards characterizing \sol\ deductively we focus on a
definition of program semantics within pure second-order logic based
on relational quantification.
This approach has the dual advantage of
facilitating a proof-theoretic analysis while supporting ontological considerations.

Let \sms{\calL_2} be the language of second-order logic (\sol) 
based on relational quantifiers, as follows.
We posit variables \sms{R_i^k} (\mm{i,k \geq 1})
intended to range over \sm{k}-ary relations. 
The choice of relations rather than functions or
sets is generally inconsequential \cite[\S 15]{Avigad23}, but 
quantifying over relations presents conceptual as well as technical advantages.
The set of atomic-formulas is augmented with phrases of the form 
\SM{R_i^k(\bmt_1,\ldots, \bmt_k)}, and the formation-rules for formulas
are augmented with relational quantification: if \sms{\grf}
is a formula then so is \sms{\forall\, R^k \, \grf}. 
For the sake of expository economy 
we don't, without loss of generality, use existential relational-variables.

\subsection{Formalization of program semantics}
We expand the expressive machinery of first-order logic
to allow relational quantifiers, while avoiding the
proof-theoretic power of second-order logic by
restricting the Comprehension Principle to first-order
predicates. 
The choice of relations, rather than functions or sets,
has the dual advantage of
facilitating a proof-theoretic analysis while supporting 
ontological considerations.

Posit variables \sms{R_i^k} (\mm{i,k \geq 1})
intended to range over \sm{k}-ary relations. 
The set of atomic-formulas is augmented with phrases of the form 
\SM{R_i^k(\bmt_1,\ldots, \bmt_k)}, and the formation-rules for formulas
are augmented with relational quantification: if \sms{\grf}
is a formula then so is \sms{\forall^2 R^k \, \grf}. 
(W.l.o.g. we restrict attention to the universal relational-quantifier.)


\bdfnl{dfn:M}
Given a \vv-program \sms{\gra} using variables \sms{x_1 \ldots x_k}
we define a second-order \vv-formula
\sms{M_\alpha} whose free variables include
\sms{x_1,\ldots,x_k,y_1,\ldots,y_k}.
The formulas \sms{M_\alpha} are defined by recurrence on \sm{\gra}
by cases on the main construct.
We write \sms{M_\alpha[\vec{u},\vec{v}]} 
for \sm{M_\alpha[\vec{u},\vec{v}/\vec{x},\vec{y}]}.

%
\ph\bit{Assignment:} \sms{\gra} is \SM{x_{i_1},\ldots,x_{i_k}\, := \, \bmt_1,\ldots,\bmt_k}.
	\ms{M_\alpha} is 
		\SM{\bigwedge_{j=1\ldots k}\, y_{i_j} \seq \bmt_j
		\; \wedge \; \bigwedge_{j\neq i_1\ldots i_k}\, y_j \seq x_j}.\\[1mm]
%
%
\ph\bit{Test:}  \sms{?\,\grf}.
	\ms{M_\alpha} is 
	\SM{\grf \wedge \vec{y} \seq \vec{x}}.\\[1mm]
\ph\bit{Composition:}  \sms{\grb \comp \grg}.
	\SM{M_\alpha} is 
		\SM{\exists\, \vec{z} \; M_\beta[\vec{x},\vec{z}] \; \wedge \; 
				M_\gamma[\vec{z},\vec{y}]}.\\[1mm]
\ph\bit{Union:}  \sms{\grb \, \pmb{\ccup} \, \grg}.
	\sms{M_\alpha} is 
	\SM{M_\beta \; \vee \; M_\gamma}.\\[1mm]
\ph\bit{Iteration:} \sms{\grb^\star}.\\
	\sms{M_\alpha} is \QM{\forall \vec{x},\vec{y} 
		\, \forall R^k \, \cl_\beta[R] \ra (R(\vec{x}) \ra R(\vec{y})}\\
where \SMS{\cl_\beta[R]} abbreviates
\QM{\forall\, \vec{u},\vec{v}\; R(\vec{u}) \wedge M_{\beta}[\vec{u},\vec{v}] 
		\ra R(\vec{v})}.

\edfn



\bthml{thm:M}
The formulas \sms{M_\alpha} defines correctly the semantics of programs.
That is, for all \sms{\vec{a},\vec{b} \in |\calS|^k} we have
\SMS{[\vec{a}/\vec{x}] \! \ara{\alpha} \! [\vec{b}/\vec{x}]} iff
\SMS{\calS, [\vec{a},\vec{b}\,/\,\vec{x},\vec{y}] \models M_{\alpha}}
\ethm
The proof is by a straightforward induction on \sm{\gra}.

\vspace{-4mm}


\bdfn
For a \vv-pca \sm{\grp} of the form \bxrm{\sm{\grf \, \plbk \,\gra\,\prbk\,\grq}}
we write \sms{\hat{\grp}} for the formula that conveys \sm{\grp} in second-order logic,
namely
\QM{\forall\, \vec{x},\vec{v},\vec{z} \, (
	\grf \wedge\; M_\alpha[\vec{x},\vec{v}]
		\,\ra\, \grq[\vec{v}/\vec{x}]})
\edfn
where \sms{\vec{z}} are the variables free in \sms{\grf,\grq} but not in \sms{\gra}

\subsection{Henkin models}

A naive definition of truth for second-order formulas is analogous to
that for first-order formulas.
The truth of first-order formulas
in a structure \sms{\calS} is defined modulo valuations \sms{\grh} 
of free-variables as structure elements.
Likewise, the truth of second-order formulas in \sms{\calS} is defined
modulo a valuation as above, as well as a valuation
\sms{H} 
of relational-variables as relations over \sm{\calS}.
A formula \MS{\forall^2\, R^k\; \grf} is then true in \sms{\calS} 
modulo \sms{\grh,H} iff \sms{\grf} is true in \sms{\calS} 
modulo \sms{\grh,H'} for every extension
\sms{H'} of \sms{H} that encompasses \sm{R}.

A challenge appears as soon as we attempt to axiomatize
second-order logic by a deductive system \bxsf{L}$_2$.
Of course, we'll promise to only prove
formulas that are valid for the ``standard interpretation'' above.
Since the set of valid second-order formulas
is not definable in second-order arithmetic, let alone RE,
any such proof-system would only yield
some, but not all, valid second-order formulas.
The semantics that \bxsf{L_2} would capture must therefore
allow ``structures'' that falsify formulas that are true in the
standard interpretation above, but not provable in \bxsf{L}$_2$.
Henkin's insight \cite{Henkin50} was to consider a non-standard
notion of modeling, in which relational variables range over
a prescribed family of relations which is part of the model.
The coherence of this construction
requires that those families of relations satisfy certain
closure conditions.
The interested reader is advised to consult \cite{Henkin50} for detail.

\subsection{A sequential calculus for \flp}

We focus on a sequential calculus for \flp\ in the style of 
\cite[Chapter 3]{Takeuti75}.
The sequential style is preferred, since it is
particularly handy for tracking formulas' polarity.

The sequential rules are as follows (using common conventions 
for displaying sequents,
such as \sms{\grG,\grf} standing for \SM{\grG \cup \{\grf\}},
with \sms{\grf \in \grG} allowed).

\noindent

Let \sms{\grf[\grl u_1 \cdots u_n \grq/R]} 
stand for the formula obtained by replacing
in \sms{\grf} each subformula \sms{R(\bmt_1,\ldots,\bmt_n)} 
by \SM{\grq[\bmt_1 \ldots \bmt_n/u_1 \ldots u_n]}.
(The variables \sms{\vec{u}} are 
merely place-holders for the substitution, and are assumed 
to not occur in \sm{\grf}.)
We refer to phrases of the form \sms{\grl u_1 \cdots u_n \grf} 
as \bit{predicates}, and dub variables other than \sms{\vec{u}}
occurring free in \sms{\grf} \bit{side-variables}.

The inference rules for sequents are commonplace \cite{Takeuti75})
Of interest here are the two rules for universal quantifier over relations:
$$
\begin{array}{ll}
\begin{array}{c}
\textsf{Antecedent Rule}\\[2mm]
\infer{\grG, \, \forall R \grf \; \rA \; \grD}
		{\grG, \, \grf[\grl \vec{u}\grq/R], \, \grG \, \rA \, \grD}\\[1mm]
		\hbox{(\ms{\grq} a \vv-formula)}
\end{array}
\qquad  & \qquad 
\begin{array}{c}
\textsf{Succedent Rule}\\[2mm]
\infer{\grG \; \rA \; \forall R \, \grf, \;\grD} 
			{\grG \; \rA \; \grf, \, \grD}\\[1mm]
	\hbox{(no free \sms{R} in conclusion)}
\end{array}
\end{array}
$$

The antecedent-rule conveys the Principle of Comprehension,
which is also rendered by the schema
$$
\forall \vec{x} \; \exists R \; \forall \vec{z} \; R(\vec{x},\vec{z}) \; \lra \; \grf
$$
where \sms{\grf} is a formula with free variables \sm{\vec{x}}.
We assume that the Cut Rule is not included, 
since it is eliminable \cite{Girard71}.

i
\subsection{Proofs of \flp}

The huge deductive power of the calculus above for second-order logic resides
in the scope of relational quantifiers.
Recall that the definition of
\SMS{\calS,\,\grh,\,H \; \models \; \forall R^k \, \grf}
requires \SMS{\calS,\grh, H\cdot [A/R] \models \grf} for all \sm{A \subset |\calS|^k}. 
But the inference rules might remain correct even if \sm{A} ranges 
over a sub-collection 
\sms{\calC} of \sm{\calP(|\calS|^k)}, 
a possibility envisioned by Henkin \cite{Henkin50}.
The Comprehension principle requires that \sms{\calC} include all the
relations definable by second-order formulas, and the relational quantifiers in such 
definitions are themselves interpreted with respect to \sms{\calC}.
In short, \sms{\calC} must be closed under definitions by second-order formulas.
This circular situation may be unsettling, but it disappears once we weaken
Comprehension to relations that are \bit{first}-order definable.


We thus focus on a variant \flp\ of \fol, 
in which all predicates in Comprehension are defined by first-order formulas. 

A sequential calculus for \flp\ is thus simply the calculus
for \sol\ with the stipulation that the eigen-formula of the 
\sm{\forall^2} Antecedent Rule 
be a first-order \vv-formula.

The main take-away here is that \esl\ has no ontological or 
epistemic premises beyond
the admission of first-order definitions of relations.
\esl\ is analogous to the subsystem \sms{\bxsf{ACA}_0}
of second-order Arithmetic \cite{Simpson09}, but it is not the same!
\esl\ is an uninterpreted logic, with a generic notion of semantics,
whereas \sms{\bxsf{ACA}_0} is a second-order theory for the
particular structure of natural numbers with a collection of basic operations.
We allow free relational-variables (i.e.\ parameters)
in eigen-predicates, as does \SM{\bxsf{ACA}_0}.
This is innocuous for both \esl\ and \SM{\bxsf{ACA}_0},
but the theory \sms{\bxsf{ACA}_0} includes, in addition, a template of Induction, 
and letting induction-formulas have free relational variables \bit{does} make a
difference.  Evidently, we need not worry about this here.

\section{Soundness of \hl\ for \flp}

Despite its limitations \flp\ suffices to capture \hl:

\bthml{thm:soundness}
\zero \bxrm{(Deductive soundness of \hl)}
If a pca \sms{\grp} is provable in \hl\
then \sms{\hat{\grp}} is provable in \flp.
\ethm

\prf
The proof is by induction on derivations \sms{\calD} of \hl,
inspecting cases for the final inference of \sm{\calD}.\\[2mm]
\ph \bit{Assignment:} \ms{\calD} consists of a single pca \sms{\grp} of the form
	\SM{\grf[\vec{\bmt}/\vec{x}] \;\plbk \vec{x} \! := \! \vec{\bmt} \prbk\; \grf}.\\
\ms{\hat{\grp}} is
\SM{(\grf[\vec{\bmt}/\vec{x}])[\vec{u}/\vec{x}] \; 
	\wedge \; \vec{v} =\vec{\bmt}[\vec{u}/\vec{x}])
	\; \wedge \;
	  (\bigwedge_{j \neq i}\, v_j \seq u_j) \; \ra \; \grf[\vec{v}/\vec{x}])}.
This is provable from the equational rule of first-order logic.\\[2mm]
\ph \bit{Test:} 
\ms{\calD} consists of a single pca \sms{\grp} of the form
\SMS{\grf \, \plbk ?\, \grx \, \prbk \grq} where
\SMS{\grf \wedge \grx \ra \grq} is provable in \fol.
\MS{\hat{\grp}} is thus 
\SMS{\grf[\vec{u}/\vec{x}] 
	\; \wedge \; (\grx[\vec{u}/\vec{x}] \wedge \vec{u}= \vec{v})
	\;  \ra \; \grq[\vec{v}/\vec{x}]}
which follows in \fol\ from \SM{\grf \wedge \grx \ra \grq}.\\[2mm]
\ph \bit{Composition:}
\ms{\calD} infers a pca \sms{\grf\, \plbk\, \grb\comp\grg\, \prbk \; \grq}
from pca's \SMS{\grf \; \plbk \, \grb \, \prbk \, \grch} and
			\SM{\grch\; \plbk \, \grg \, \prbk \; \grq}.
By IH 
\SMS{\grf[\vec{u}/\vec{x}] \, \wedge \, M_{\beta}[\vec{u},\vec{w}] 
	\ra \grch[\vec{w}]} 
and 
\SMS{\grch[\vec{w}] \wedge M_{\gamma}[\vec{w},\vec{v}] \ra \grq[\vec{v}]}
are provable in \flp.\\
So 
\SM{\grf[\vec{u}/\vec{x}] \, \wedge \, 
	\exists \vec{w} \;(\,M_{\beta}[\vec{u},\vec{w}] 
		\; \wedge \; M_{\gamma}[\vec{w},\vec{v}])
	\; \ra \; \grq[\vec{v}/\vec{x}]},\\
 i.e.\
\SMS{\grf[\vec{u}/\vec{x}] \, \wedge \,
	M_{\beta\! \comp\i! \gamma}[\vec{u},\vec{v}] 
	\; \ra \; \grq[\vec{v}/\vec{x}]}
is provable in \flp.\\[2mm]
\ph \bit{Union:}
\ms{\calD} infers a pca \SMS{\grf\; \plbk\, \grb\ccup\grg\, \prbk \; \grq}
from the two pca's \SMS{\grf \; \plbk \, \grb \prbk \, \grq} and
\SM{\grf \; \plbk \, \grg \prbk \, \grq}.
By IH 
\SMS{\grf[\vec{u}/\vec{x}] \, \wedge \, M_{\beta}[\vec{u},\vec{v}] 
	\ra \grq[\vec{v}/\vec{x}]} 
and 
\SMS{\grf[\vec{u}/\vec{x}] \wedge M_{\gamma}[\vec{u},\vec{v}] 
		\; \ra \; \grq[\vec{v}/\vec{x}]}
are provable in \flp, from which the provability of
\SMS{\grch[\vec{u}] \wedge M_{\beta \ccup \gamma}[\vec{u},\vec{v}] 
	\ra \grq[\vec{v}]}, that is \sm{\hat{\grp}},
follows trivially.\\[2mm]
\ph \bit{Iteration:}
\ms{\calD} infers a pca \SMS{\grf\; \plbk\, \grb^*\, \prbk \; \grf}
from \SM{\grf \; \plbk \, \grb \prbk \, \grf}.
By IH
\SMS{\grf[\vec{u}/\vec{x}] \wedge M_{\beta}[\vec{u},\vec{v}] 
	\;\, \ra \; \grf[\vec{v}/\vec{x}]}
is provable in \flp.
This is precisely \SM{\cl_{\beta}[\grl \vec{u}.\grf]}.
So \sms{\hat{\grp}}, namely
\begin{center}
$\grf[\vec{u}/\vec{x}]
  \, \wedge \, (\forall R \; \cl_{\beta}[R]\, 
		\wedge \, R(\vec{u}) \ra R(\vec{v})) 
			\;\, \ra \; \grf[\vec{v}/\vec{x}]$
\end{center}
is provable in \flp\ 
by instantiating \SMS{\forall R} in the premiss
by the predicate \SM{\grl \vec{x}.\grf}.\\[1mm]
\ph \bit{Consequence:}  Trivial.\qed

\section{Completeness of \hl\ for \flp}

We prove here our main result:

\bthml{thm:completeness}
\zero\bxrm{(Deductive completeness of \hl)}
Given a pca \sms{\grp}, if \sms{\hat{\grp}} is provable in \flp\ 
then \sms{\grp} is provable in \hl.
\ethm

The proof proceeds by induction on \sms{\gra}, using
technical proof-theoretic tools presented in the next two subsections.

\subsection{Polarity of $\forall^2$}

\medskip


\bdfn
An occurrence of a phrase (term, subformula or quantifier) 
in a formula \sms{\grf}
is \bit{positive (negative)} in \sms{\grf} if it is in the negative scope
of an even (odd, respectively) number of implications, 
where a negation \sms{\neg\grq} is defined as \SM{\grq \sra \bot}. 
Such a phrase-occurrence in
\sms{\grG \rA \grD} is \bit{positive (negative)} in the sequent  
if it is a positive (negative) occurrence in \sms{\grG}
or a negative (positive) one in \sm{\grD}.

A formula \ms{\grf} is 
\bit{\sm{\forall^2}-positive (\sm{\forall^2}-negative)}
if it has no negative (positive) occurrence of \sm{\forall^2}, and
a sequent \SMS{\grG \rA \grD} is \bit{\sm{\forall^2}-positive} 
if every formula in \sms{\grG} is \sm{\forall^2}-negative and every formula in 
\sms{\grD} is \sm{\forall^2}-positive.
\edfn

From the definition of \sm{M_{\alpha}} (Def \ref{dfn:M}) we obtain

\bleml{lem::Mpositive}
For every program \sms{\gra} the formula \SMS{M_{\alpha}} is \sm{\forall^2}-positive.
Consequently, for any pca \sms{\grp} the second-order formula 
\sms{\hat{\grp}} is \sm{\forall^2}-negative. \qed
\elem

\prf  By induction on \sm{\gra} and cases for its main construct.

\bi
\ti If \sms{\gra} is an assignment or a test then \sms{M_\alpha} is
a first-order formula, which is positive trivially.
\ti If \sms{\gra} is \SMS{\grb \comp \grg} then
        \SMS{M_\alpha[\vec{u},\vec{v}]} is
                \SM{\exists \vec{z} \; M_\beta[\vec{z}/\vec{v}] \; \wedge \;
                                M_\gamma[\vec{z},\vec{u}]}
with all occurrences of \sms{\forall^2} in \SMS{M_\beta} and \SMS{M_\gamma} 
positive in these formulas.  Since the latter are positive subformulas
of \sms{M_\alpha}, all occurrences of \sms{\forall^2} in \sms{M_\alpha}
are positive.
\ti The case for Union is similar.
\ti If \sms{\gra} is \SMS{\grb^\star} then
        \SMS{M_\alpha[\vec{u},\vec{v}]} is\\
        \SMS{\forall R \; \cl_\beta[R] \wedge R(\vec{u}) \ra R(\vec{v})}
        where \sms{R} is a relational variable.

\medskip

The occurrences of \sms{\forall^2} in \sms{M_\alpha} are
those in \sms{M_\beta} as well as the main \sms{\forall R}.
The latter is trivially positive in \sms{M_\alpha}, whereas the
ones in \sms{M_\beta} are negative in \sms{\cl_\beta},
which in turn is negative in \sms{M_\beta}, so they are all
positive in \sm{M_\alpha}.\qed
\ei

The following observation motivates our definition of quantifier-expansion
below.

\bleml{lem:cutfree-polarity}
If a derivation \sms{\calD} of \flp\ derives a \sm{\forall^2}-negative sequent,
then all sequents in \sms{\calD} are \sm{\forall^2}-negative.
In particular, there is no \sms{\forall^2}-succedent inference in \sm{\calD}.
\elem

\prf The proof is by a straightforward induction on sequential derivations.
(Recall that our proofs are cut-free.) \qed

%
%
%


\subsection{Expansion of negative \sm{\forall^2}}

In a derivation \sms{\calD} of \flp\ the rule \SMS{\forall^2 L} 
is used a finite number of times, say with eigen \vv-predicates
\sms{\grl \vec{u}. \grx_i} ($i= 1\ldots n$), all of arity \sm{k}, and with
side-variables  \SMS{\vec{v}_i} respectively.
This suggests that for \calD\ to make sense it suffices, perhaps, to
admit a universe of relations that consists just of the predicates above.
If so, then every formula \SMS{\forall R \, \grf}
might be replaced  by \SM{\wedge_{i=1\ldots n} \grf[\grl \vec{u}. \grx_i/R]}.
This might be done hereditarily, since the replacing above, with
\sms{\grf} first-order, does not introduce any new occurrences of 
\SMS{\forall^2 R} inferences.

A potential concern with the replacement of \sms{\forall R\, \grf}
by a long conjunction is that the presence of 
the predicates' side-variables \sms{\vec{u}} might invalidate
first-order inferences elsewhere in \sms{\calD}.
This concern is addressed by
refining the long conjunction above to 
	\SM{\wedge_{i=1\ldots n}
		\forall \vec{v}_i \; \grf[\grl \vec{u}. \grx_i/R]}.

The replacement by a conjunction might conceivably
invalidate instances of \sms{\forall^2} elsewhere in \sm{\calD}
that close the variable \sms{R}.
But if \sms{\calD} derives a \sms{\forall^2}-negative 
sequent then Lemma \ref{lem:cutfree-polarity} ensures that no such 
instances can occur!
 

\bdfn
Let \SMS{\grX} be a finite set \SMS{\lb\grl \vec{u}.\grx_i \rb_i} of 
predicates of common arity.
The \bit{\grX-expansion} of a negative sequent \sm{\grs},
denoted \sm{\grs^{\Xi}}, is obtained by replacing in \sms{\grs} 
each subformula \SMS{\forall^2\, R \; \grq[R]} by 

\begin{center}
\mm{\bigwedge_i \; \grq^{\Xi}[\grl \vec{u}\; \grx_i[\vec{x},\vec{u}]]}\\[-5mm]\qed
\end{center}
\edfn

We consider the sequential calculus for \fol\ obtained from
\flp\ by omitting the \sms{\forall^2} rules.

\bleml{lem:expansion-out}
Let \sms{\grs} be a \sms{\forall^2}-negative sequent 
and \SMS{\Xi} a set of (first-order) predicates.
If \sms{\grs^\Xi} has a \fol-derivation then
\sm{\grs} is provable in \flp.
\elem

\prf For any first-order formula \SMS{\grf[R]} with one free relational-variable
\sms{R} the sequent \SMS{\forall R \grf[R] \; \rA \; \grf^\Xi}
is trivially derivable in \flp. Since our proof system is closed under Cut,
	having \SMS{\grG, \, \grf^\Xi[R] \; \rA \; \grD} derivable in \fol\
implies that \SMS{\grG, \, \forall R \grf[R] \; \rA \; \grD} is derivable in \flp.

The Lemma follows
by main induction on the total number of 
logical-operations in \sms{\grs} with \sms{\forall^2} in their scope,
secondary induction on the number of occurrences of \sms{\forall^2}
	in \sm{\grs}, and tertiary induction on \calD.\qed

%
%
%
%

A converse of Lemma \ref{lem:expansion-out} is

\bleml{lem::eigen-expansion}
If \sms{\calD} is a derivation in \flp\ of a negative sequent \sms{\grs}
then the sequent \SMS{\grs^{\Xi}} is derivable in \fol,
where \sms{\grX} consists of all eigen-predicates of \sms{\forall^2 L} inferences
in \sm{\calD}.
\elem

\prf Straight-forward induction on \sm{\calD}.\qed

The expansion construction presented above is inspired by the proof in
\cite{Troelstra73} that arithmetic comprehension is conservative 
over First-Order Arithmetic.

\subsection{\hl\ is complete for \flp}

\bthml{thm::completeness}
\zero \bxrm{(Deductive completeness of \hl)}
For every pca \sm{\grp}, if \sms{\hat{\grp}} is provable in \flp\
then \sms{\grp} is provable in \hl.
\ethm

\prf
The proof is by induction on \sm{\gra}. 

\ph \bit{Assignment:}
	\ms{M_\alpha} is
	\sm{\vec{y} \seq \vec{\bmt}}, so \sms{\hat{\grp}}
is \sms{\grf[\vec{u}/\vec{x}] \wedge \vec{y} \seq \vec{\bmt} \ra \grf[\vec{y}/\vec{x}]}.
which we assume to be provable in \flp.
Let \sms{\grch} be 
	\SM{\forall \vec{z} (\,\vec{z} \seq \vec{x} \ra \grf[\vec{z}/\vec{x}]}.
The formula \SMS{\grf[\vec{\bmt}/\vec{x}] \ra \grch[\vec{\bmt}/\vec{x}]} 
is then provable in \fol.
By the Assignment Rule of \hl\
\SM{\grch[\vec{\bmt}/\vec{x}]\, 
	\plbk \, \vec{x} := \vec{\bmt}\,\prbk
		\, \grch}.
Also, \SMS{\grch \ra \grf} in \fol.
Combining the three by the Consequence Rule, we get
\SM{\grf[\vec{\bmt}/\vec{x}]\, \plbk \, \vec{x} := \vec{\bmt}\, \prbk\, \grf}.
\zero\\[1.5mm]

\ph \bit{Test:} \ms{\gra} is \SMS{?\,\grx[\vec{x}]}. Suppose
\sms{\grf \wedge \xi \ra \grq}
is provable in first-order logic.
\SMS{M_\alpha[\vec{u},\vec{v}]} is 
	\SM{\grx[\vec{u}/\vec{x}] \wedge \vec{v} = \vec{u}}.
which then implies \sms{\grf[\vec{v}]} by the equational rules of \fol.\\[1mm]
\ph \bit{Composition:} \ms{\gra} is \sm{\grb \comp \grg}.
Suppose there is derivation \sms{\calD} of \flp\ of
\beqn\label{eq:comp}
\zero\qquad\qquad 
\grf[\vec{u}/\vec{x}] \ra \, (\, M_\alpha[\vec{u},\vec{v}]
	\ra \grq[\vec{v}/\vec{x}] \,)
\eeqn
By the definition of \sm{M_{\beta;\gamma}}, namely
\SM{\exists \vec{z} \; 
	M_\beta[\vec{x},\vec{z}] \; \wedge \; 
		M_\gamma[\vec{z},\vec{y}]},
(\ref{eq:comp}) implies in \fol\
$$
(\grf[\vec{u}/\vec{x}] \;\wedge \; M_\beta[\vec{u},{\vec{z}]})
	\;\; \ra \;\;
	 (M_\gamma[\vec{z},{\vec{v}]}
        \; \ra \; \grq[\vec{v}/\vec{x}])
$$

Let \sms{\grX} be the set of eigen-predicates of \mm{\forall^2}L 
inferences in \sm{\calD}.
By Lemma \ref{lem::eigen-expansion} we obtain that the first-order formula
$$
(\grf[\vec{u}/\vec{x}]\; \wedge \; M_{\beta}[\vec{u},\vec{z}]^\Xi) \;\; \ra \;\;
	(M_{\gamma}[\vec{z},\vec{v}]^\Xi  \; \ra \;  \; \grq[\vec{v}/\vec{x}])
$$
is provable in \fol.
The premise of this implication has no occurrence of \sm{\vec{v}},
and its conclusion has no occurrence of \sm{\vec{u}}.
So by the Interpolation Theorem for \fol\ there is a 
first-order formula \sm{\chi}, with no free occurrences of
either \sms{\vec{u}} or \sm{\vec{v}}, such that
$$
\grf[\vec{u}/\vec{x}]\, \wedge \,
        M_{\beta}[\vec{u},\vec{z}]^\Xi) \,\;\ra \; \chi
\qquad \text{and} \qquad
\chi  \,\wedge \, M_{\gamma}[\vec{z},\vec{v}]^{\Xi}
	\;\, \ra \; \grq[\vec{v}/\vec{x}]
$$
are both provable in \fol.

By Lemma \ref{lem:expansion-out} this implies that 
$$
\grf[\vec{u}/\vec{x}]\; \wedge \;
        M_{\beta}[\vec{u},\vec{z}] \; \ra \; \chi
\qquad \text{and} \qquad
\chi \wedge M_{\gamma}[\vec{z},\vec{v}]
$$
are provable in \flp, which by induction assumption implies that
\sms{\grf \; \plbk \grb \prbk \; \grch}
and
\sms{\grch \; \plbk \grg \prbk \; \psi} are both provable in \hl.
Applying the Composition Rule of \hl\ we conclude that
\SMS{\grf \; \plbk \grb \comp \grg \prbk \; \psi} is likewise provable.\\[1mm]
%
%
%
%
%

\ph \bit{Union:} \ms{\gra} is \SM{\grb \ccup \grg}.
Suppose there is derivation \sms{\calD} of \flp\ of
\begin{equation}\label{eq:branch}
\zero\qquad\qquad 
	\hbox{$\grf[\vec{u}/\vec{x}] \; \wedge \; M_\alpha[\vec{u},{\vec{v}]}
	\;\;\ra \;\; \grq[\vec{v}/\vec{x}]$}
\end{equation}
By the definition of \sms{M_\alpha} in terms of
	\SMS{M_\beta} and \SM{M_\gamma}, namely
\SMS{M_\beta[\vec{u},\vec{v}] \; \vee \; 
		M_\gamma[\vec{u},\vec{v}]}
(\ref{eq:branch}) implies in \fol\
$$
\grf[\vec{u}/\vec{x}] \;\wedge \; M_\beta[\vec{u},{\vec{z}]})
        \;\, \ra \;\; \grq[\vec{v}/\vec{x}]) 
	\qquad \text{and} \qquad 
\grf[\vec{u}/\vec{x}] \;\wedge \; M_\gamma[\vec{u},{\vec{z}]})
        \;\, \ra \; \grq[\vec{v}/\vec{x}]
$$
By induction assumption these imply that
\SMS{\grf \; \plbk \grb \prbk \; \grq}
and
\SMS{\grch \; \plbk \grg \prbk \; \grq} are both provable in \hl.
By the Union Rule of \hl\ we obtain
\SM{\grch \; \plbk \grb \ccup \grg \prbk \; \grq}.\\[2mm]

%
%
%
%
\ph \bit{Iteration:}
\ms{\gra} is \SMS{\grb^*}.
Suppose \sms{\grp} is \SMS{\grf \; \plbk \grb^* \prbk \; \grq},
and \SMS{\calD} is a cut-free derivation of \flp\ for \sms{\hat{\grp}}.
i.e.\ for
$$
\grf[\vec{u}] \; \wedge \; (\forall R  \; R(\vec{u})  \;
        \ra \cl_{\beta}[R] \ra R(\vec{v})) \quad \ra \quad \grq[\vec{v}]
$$
where \SMS{\cl_{\beta}[R]} is
$$
\forall \vec{w},\vec{z}\; R(\vec{w}) \wedge 
		M_\beta[\vec{w},\vec{z}] \ra R(\vec{z})
$$
Let \sms{\grX} be the set of eigen-predicates in \sm{\calD}, as for Composition above. 
Then, by Lemma \ref{lem::eigen-expansion},

\begin{equation}\label{eq:1}
\zero
\hbox{$\grf[\vec{u}] \; \wedge \;  (\bigwedge_i  \; \grx_i[\vec{u}]  \;
        \ra \cl_{\beta}[\grx_i]^{\Xi} 
			\ra \xi_i[\vec{v}]) \;\;\ra \;\; \grq[\vec{v}]$}
\end{equation}
is provable in \fol.

Let \sms{\grch[\vec{p}]} stand for
$ (\bigwedge_i\; \xi_i[\vec{p}] \ra 
	\cl_{\beta}[\grx_i]^{\Xi} \ra \grx_i[\vec{v}])  
	\;\;\ra \;\; \grq[\vec{v}]$.
Then\\[2mm]
\bi
\ti The first-order formula 
\begin{equation}\label{eq0}
{\grf[\vec{u}] \ra \grch[\vec{u}]}
\end{equation}
paraphrases \ref{eq:1}, and is thus provable in \fol.
 
\ti The formula
\begin{equation}\label{eq1}
\grch[\vec{p}/\vec{x}] \; \wedge \; M_{\beta}[\vec{p},\vec{q}] 
	\;\; \ra \; \grch[\vec{q}.\vec{x}]
\end{equation}
is easily provable in \esl.

\ti  For each \sms{\grx_i} the implication
\begin{equation}\label{eq2}
\xi_i[\vec{p}/\vec{x}] \;\; \ra \;\;
        (\cl_{\beta}[\grx_i]^{\Xi} \;\;\ra \; \grx_i[\vec{v}/\vec{x}]))
\end{equation}
it trivial where \sms{\vec{p}} is \sm{\vec{v}}.
So the premiss of \SMS{\grch[\vec{v}]} is equally trivial, implying that
\SMS{\grch[\vec{v}] \;\ra \; \grq[\vec{v}]} is provable in \fol.
\ei

By induction assumption the provability in \esl\ of (\ref{eq1}) implies 
that \SMS{\grch\,\plbk\, \grb\,\prbk\,\grch} is provable in \hl. 
From this it follows, by the Iteration Rule, that
\SMS{\grch\,\plbk\, \grb^* \, \prbk\,\grch} is provable in \hl.

From the provability in \fol\ of (\ref{eq0}) and (\ref{eq2}) we obtain,
by applying twice the Consequence Rule of \hl,
that \SMS{\grf\,\plbk \, \grb^* \, \prbk\,\grq} 
is provable in \hl\ as well.\qed

\section{Characterization theorems}

\subsection{\esl\ characterizes \hl}
\bthml{thm:delineation}
\zero \bxrm{(Deductive characterization of \hl)}
A pca \sms{\grp} is provable in Hoare Logic iff its explicit rendition
\sms{\hat{\grp}} is provable in \esl.

The same equivalence holds for the constructive (i.e.\ intuitionistic) versions
of \hl\ and \esl.
\ethm

\prf The first equivalence combines
Theorems \ref{thm:soundness} and \ref{thm:completeness}.

To see that the equivalence holds for the constructive versions of
\hl\ and \esl\ observe that the proofs of Theorems
\ref{thm:soundness} and \ref{thm:completeness} remain correct once
the Consequence Rule of \hl\ is stated for provability in intuitionistic first-order 
logic, and the rules of \esl\ are restricted to sequents with at most
one formula in the succedent (i.e.\ the sequents' right-hand side).\qed

\medskip

The result for constructive logics illustrates one advantage of
proof-theoretic methods over model-theoretic methods, which are
far more complex to adapt to constructive formalisms. 

Theorem \ref{thm:delineation} identifies \hl\ as capturing
first-order reasoning about regular programs, since \esl\
does not go beyond an acceptance of first-order definitions,
which already underlies  first-order logic itself.

Note that the essence of the theorem is not tied to the
specific second-order definition we adopt here for program
semantics.  That second-order definition is correct for
standard semantics no matter our focus on \esl.

\subsection{Characterizations modulo a theory}
\bdfn Let \sms{\bmT} be a \vv-\bit{theory}, i.e.\ a 
set of \vv-formulas.
\bit{Hoare Logic modulo \sm{\bmT}} is the deductive formalism obtained 
from \hl\ by replacing the provability 
premises of the Consequence Rule
by provability from \sm{\bmT}.\qed
\edfn

\bthml{thm:charact-modulo}
A pca \sms{\grp} is provable in
\hl\ modulo a \vv-theory \sms{\bmT} iff its explicit rendition 
\sms{\hat{\grp}} is provable in \esl\ from \sm{\bmT}.
\ethm

\prf Repeat mutatis mutandis the proofs of Theorems 
\ref{thm:soundness} and \ref{thm:completeness}.\qed

\subsection{Semantic completeness for Henkin's models}

Consider Henkin's \vv-models for \sol\ as defined in \cite{Henkin50},
with simple adaptation to our context. 
The basic idea is that relations over objects are treated 
as structure elements of separate sorts, one for each arity, whereas
the basic objects fall under their own sort.
The sorts are tied together by a binary relation \sms{\gre} 
that holds between
\sm{k}-ary tuples of ``object''-elements and elements
that are ``\ms{k}-ary relations''.
This is akin to treating second-order logic as a
multi-sorted first-order theory.


We obtain then from Theorem \ref{thm:delineation}

\bthml{thm:henkin}\zero \bxrm{(Henkin-model 
	characterization of \hl)}
A pca \sms{\grp} is provable in Hoare Logic iff 
it is true in every Henkin model of \esl.
\ethm

\newpage

\bibliography{x}

\end{document}

%% file: symbol.tex
\usepackage{amsfonts}
\usepackage{relsize}
\usepackage{stackengine}
\usepackage{amssymb}
\usepackage{amscd}
\usepackage{tikz,lipsum,lmodern}

\newcommand{\lb}{\{} 
\newcommand{\rb}{\}} 

\newcommand{\plbk}{\hbox{\pmb{\textbf{[}}}}
\newcommand{\prbk}{\hbox{\pmb{\textbf{]}}}}

\newcommand{\ra}{\rightarrow}
\newcommand{\pa}{\rightharpoonup}
\newcommand{\sra}{\!\rightarrow\!} 

\newcommand{\rA}{\Rightarrow}  

\newcommand{\lra}{\leftrightarrow}

\newcommand{\seq}{\!=\!}

\newcommand{\comp}{{\textrm{\pmb{\onemm;\onemm}}}}

\newcommand{\ara}[1]{\xrightarrow{#1}} 

%% file: unit.tex
\newtheorem{prop}{Proposition}
\newtheorem{thm}[prop]{Theorem}
\newtheorem{lem}[prop]{Lemma}
\newtheorem{cor}[prop]{Corollary}
\newtheorem{dfn}[prop]{Definition}

\newcommand{\prf}{{\sc Proof.\onemm}}

\newcommand{\ti}{
	\item[]{\large \raisebox{2pt}{{\onemm\normalsize{$\vartriangleright\;\;$}}}}}
\renewcommand{\ti}{
	\item[]{\raisebox{1.6pt}{{\normalsize{$\vartriangleright\;\;$}}}}}

\renewcommand{\ti}{
	\item[]{\raisebox{-1pt}{\onemm\small{$\vartriangleright\;$}}}}

\renewcommand{\ti}{
	\item[]{\onemm\small{$\vartriangleright\;$}}}

\newcommand{\ph}{\noindent\zero\onemm}	

\newcommand{\bthm}{\begin{thm}}
\newcommand{\bthml}[1]{\begin{thm}\label{#1}}  
\newcommand{\bdfnl}[1]{\begin{dfn}\label{#1}}  
\newcommand{\bleml}[1]{\begin{lem}\label{#1}}  
\newcommand{\ethm}{\end{thm}\text}
\newcommand{\bprop}{\begin{prop}\textsl}
\newcommand{\eprop}{\end{prop}\text}
\newcommand{\blem}{\begin{lem}\textsl}
\newcommand{\elem}{\end{lem}\text}
\newcommand{\bcor}{\begin{cor}\textsl}
\newcommand{\ecor}{\end{cor}\text}
\newcommand{\bdfn}{\begin{dfn}}
\newcommand{\edfn}{\end{dfn}}

\newcommand{\bc}{\begin{center}}
\newcommand{\ec}{\end{center}}
\newcommand{\beq}{\begin{equation}}
\newcommand{\eeq}{\end{equation}}

\newcommand{\be}{\begin{enumerate}}
\newcommand{\ee}{\end{enumerate}}
\newcommand{\bi}{\begin{itemize}}
\newcommand{\ei}{\end{itemize}}
\newcommand{\bo}[1]{\begin{overlays}{#1}}
\newcommand{\eennddoo}{\end{overlays}}
\newcommand{\bd}{\begin{description}}
\newcommand{\ed}{\end{description}}
\newcommand{\beqn}{\begin{equation}}
\newcommand{\eeqn}{\end{equation}}
\newcommand{\bweqn}{\zero\\[-7mm]\begin{eqnarray*}}
\newcommand{\eweqn}{\zero\\[-14mm]\end{eqnarray*}}

\newcommand{\beqna}{\begin{eqnarray}}
\newcommand{\eeqna}{\end{eqnarray}}
\newcommand{\beqnas}{\begin{eqnarray*}}
\newcommand{\eeqnas}{\end{eqnarray*}}
\newcommand{\beqnaa}{$$\begin{array}{rcll}}  
\newcommand{\eeqnaa}{\end{array}$$}  
\newcommand{\beqnal}{$$\begin{array}{l}}  
\newcommand{\eeqnal}{\end{array}$$}  
\newcommand{\beqnana}{$$\begin{array}{lrcll}}  
\newcommand{\eeqnana}{\end{array}$$}  
\newcommand{\btbl}[1]{\begin{center}\begin{tabular}{#1}}
\newcommand{\etbl}{\end{tabular}\end{center}}
\newcommand{\beqnc}{$$\begin{array}{rclcl}}
  
\newcommand{\eeqnc}{\end{array}$$}



%% file: font.tex
\newcommand{\mx}{\makebox}

\newcommand{\zero}{\mx[0mm]{}}
\newcommand{\onemm}{\mx[1mm]{}}

\newcommand{\gra}{\hbox{$\alpha$}}
\newcommand{\grb}{\hbox{$\beta$}}
\newcommand{\grg}{\hbox{$\gamma$}}

\newcommand{\gre}{\hbox{$\varepsilon$}}
\newcommand{\grh}{\hbox{$\eta$}}

\newcommand{\grl}{\hbox{$\lambda$}}

\newcommand{\grx}{\hbox{$\xi$}}

\newcommand{\grp}{\hbox{$\pi$}}

\newcommand{\grs}{\hbox{$\sigma$}}

\newcommand{\grf}{\hbox{$\varphi$}}
\newcommand{\grch}{\hbox{$\chi$}}
\renewcommand{\grq}{\hbox{$\psi$}}  

\newcommand{\grG}{\hbox{$\Gamma$}}
\newcommand{\grD}{\hbox{$\Delta$}}

\newcommand{\grX}{\hbox{$\Xi$}}
\newcommand{\grP}{\hbox{$\Pi$}}
\newcommand{\grS}{\hbox{$\Sigma$}}

\newcommand{\calC}{\hbox{$\cal C$}}
\newcommand{\calD}{\hbox{$\cal D$}}

\newcommand{\calL}{\hbox{$\cal L$}}

\newcommand{\calP}{\hbox{$\cal P$}}

\newcommand{\calS}{\hbox{$\cal S$}}

\newcommand{\smallcalS}{\hbox{{\small $\cal S$}}}

\newcommand{\scriptcalS}{\hbox{{\scriptsize $\cal S$}}}

\newcommand{\bm}[1]{{\hbox{\bfseries{\itshape{#1}}}}}

\newcommand{\bmt}{\bm{t}}

\newcommand{\bxit}[1]{\hbox{\it #1}}

\newcommand{\bxsf}[1]{\hbox{$\sf #1$}}
\newcommand{\bxrm}[1]{{\hbox{\rm #1}}}

\newcommand{\bxbf}[1]{\mbox{\bf #1}}

\newcommand{\bft}{\hbox{\bf t}}

\newcommand{\ttf}{\hbox{$\tt f$}}

\renewcommand{\ttf}{\hbox{\tt f}}

\newcommand{\bmT}{{\bm{T}}}

\newcommand{\ttQ}{\hbox{$\tt Q$}}

\newcommand{\bit}[1]{{\textbf{\textit{#1}}}}	



%% file: mathenv.tex
\newcommand{\xx}[1]{#1}
\newcommand{\mm}[1]{\xx{$#1$}}		
\newcommand{\sms}[1]{\xx{$\, #1 \,$}}		
\newcommand{\SMS}[1]{\xx{$\,\, #1 \,\,$}}	

\newcommand{\sm}[1]{\xx{$\, #1 $}}		
\newcommand{\ms}[1]{\xx{$ #1 \, $}}		
\newcommand{\SM}[1]{\xx{$\,\, #1 $}}		
\newcommand{\MS}[1]{\xx{$ #1 \,\, $}}		
\newcommand{\QM}[1]{\xx{$\quad #1 $}}		

%% file: color.tex
\usepackage{xcolor}
\usepackage{color}
\usepackage{colordvi}

\usepackage[most]{tcolorbox}